\documentstyle[eqsecnum,aps,prd,epsf,subfigure]{revtex}
 \def\scri{\hbox{${\cal J}$\kern -.63em {\raise
      .53ex\hbox{$\scriptscriptstyle (\ $}}}}

\begin{document}
\draft

\thispagestyle{empty}
{\baselineskip0pt
\leftline{\large\baselineskip16pt\sl\vbox to0pt{\hbox{\it Department of Physics}
               \hbox{\it Kyoto University}\vss}}
\rightline{\large\baselineskip16pt\rm\vbox to20pt{\hbox{KUNS 1475}
               \hbox{April 1998}
\vss}}%
}
\vskip1cm
\begin{center}{\large \bf 
Gravitational Waves around a Naked Singularity}
\end{center}
\begin{center}{{\it - Odd-Parity Perturbation 
of Lema\^{\i}tre-Tolman-Bondi Space-Time -}}
\end{center}
\begin{center}
 {\large 
Hideo Iguchi{\footnote{e-mail: iguchi@tap.scphys.kyoto-u.ac.jp}}, 
Ken-ichi Nakao{\footnote{e-mail: nakao@tap.scphys.kyoto-u.ac.jp}},
and Tomohiro Harada{\footnote{e-mail: harada@tap.scphys.kyoto-u.ac.jp}} 
} \\
{\em Department of Physics,~Kyoto University, Kyoto 606-8502,~Japan} \\

\end{center}

\begin{abstract}
The motion of a spherical dust cloud is 
described by the Lema\^{\i}tre-Tolman-Bondi 
solution and is completely specified by initial values 
of distributions of the rest mass density and 
specific energy of the dust fluid. From generic initial conditions of
this spherically symmetric collapse, 
there appears a naked singularity at the symmetric center 
in the course of the gravitational collapse of the dust cloud. 
So this might be a counter example to the 
cosmic censorship hypothesis. 
To investigate the genericity of this example, 
we examine the stability of the `nakedness' of this singularity against 
odd-parity modes of non-spherical linear perturbations 
for the metric, i.e., linear gravitational waves.
We find that the perturbations do not diverge but 
are well-behaved even in the neighborhood of 
the central naked singularity. 
This means that the naked singularity formation process is marginally 
stable against the odd-parity modes of linear gravitational waves. 

\end{abstract}
\section{INTRODUCTION}
\label{sec:intro}

The singularity theorems revealed that the occurrence of 
singularities is a generic property of space-time in general 
relativity\cite{Penrose65,Hawking67,Hawking-Penrose70}. 
However, those theorems say nothing about the detailed features of 
the singularities themselves; for example, 
we do not get information from those theorems 
about whether the predicted singularity 
is naked or not. 
Naked means that the singularity is observable. 
A singularity is a boundary of space-time. Hence, 
in order to obtain a solution of 
hyperbolic field equations for 
matter, gauge fields and space-time itself 
in the causal future of a naked singularity, we need to impose 
a boundary condition on it. 
However, we do not yet know physically reasonable 
boundary conditions on singularities  
and hence to avoid this difficulty, 
the cosmic censorship hypotheses (CCH) 
proposed by Penrose \cite{Penrose69,Penrose79} are often 
adopted in the analysis of the physical phenomena of 
the strong gravitational fields. 

There are weak and strong versions of the CCH. 
The weak CCH states that a singularity is covered by an event
horizon and never observed by anyone included in the causal past 
of future null infinity (not globally naked) 
while the strong CCH says that 
nobody can observe a singularity (not locally naked). 
However, the validity of the CCH is 
one of the most important open questions in
classical general relativity. No one has ever proved that 
these hypotheses hold. On the contrary, 
some researchers found, analytically or numerically,
that there are solutions of the Einstein equations which have naked
singularities. If these naked singularities are physically realizable,
then we could be in an embarrassing
situation because an important assumption in theorems on the nature of a
black hole is violated. 
In the vicinity of a singularity, quantum effects 
on the space-time 
will play an important role and therefore if the existence of 
naked singularities is generic,  
in order to understand the nature of a black hole, 
we might need the knowledge of the quantum theory of gravity 
even if the black hole is a classical entity. 

In the last two decades several researchers have shown that 
the Lema\^{\i}tre-Tolman-Bondi (LTB)
space-time\cite{Tolman34,Bondi} is one of the candidates 
for a counter example to both versions of the CCH. 
This space-time describes the motion of a 
spherically symmetric inhomogeneous dust cloud and  
is completely specified by initial 
values of the rest mass density and specific 
energy of the dust fluid. 
Eardley and Smarr showed 
that the central singularity of the LTB space-time can be  
shell-focusing and naked in the case of 
marginally bound collapse\cite{Eardley}. 
Christodoulou showed that the same is true also 
for the bound case\cite{Christodoulou}. 
Newman generalized 
Christodoulou's analysis to cover a larger class of LTB 
space-times\cite{Newman}. Joshi and Dwivedi carried 
out studies of a much more 
general class of solutions in which a conical  singularity 
(but not curvature one) was allowed in the initial configuration 
and showed that the formation of a central naked
singularity is a general feature for a very wide range 
of initial data in the LTB space-time\cite{Joshi-Dwivedi}. 
These results are summarized as follows;
in this space-time, a naked singularity appears from  
generic initial data for {\it spherically symmetric} 
configurations of the rest mass density and specific energy 
of the dust fluid. 

In order to recognize this example as a serious counter 
example to the CCH, we should examine its genericity. 
That is to say, there is a possibility that the naked singularity is due to
physically unrealistic conditions, e.g., assumptions of the spherical 
symmetry, dust matter and so on. 
Shapiro and Teukolsky studied evolution of 
collisionless gas spheroids by fully general relativistic 
simulations\cite{Shapiro}. 
They found that prolate spheroids 
with sufficiently elongated initial configurations and 
even with some angular momentum, may form naked singularities.
Ori and Piran numerically examined the structure of self-similar spherical
collapse solutions for a perfect fluid with a barotropic equation of
state\cite{Ori-Piran}. They showed that there is a globally naked
singularity in a significant part of the space of self-similar
solutions. Joshi and Dwivedi analytically investigated the self-similar 
spherically symmetric collapse of a perfect fluid with a
similar equation of state\cite{JoshiII}  
and further the naked singularity produced by the gravitational 
collapse of radiation shells\cite{JoshiIII} 
and of more general matter\cite{JoshiIV}. 

In this article, we concentrate our attention on the 
issue whether the spherical symmetry is essential to 
the occurrence of the shell focusing naked singularity.  
For this purpose, we consider odd-parity modes of 
non-spherically symmetric perturbations in the 
marginally bound LTB space-time and examine the stability 
of the `nakedness' of 
that naked singularity against those linear perturbations. 
As for the non-spherically symmetric collapse case, 
Joshi and Krolack revealed that a naked singularity 
appears also in the Szekeres space-time 
with an irrotational dust matter\cite{JoshiV}.
Since the odd-parity perturbations correspond to 
the rotational motions of the dust fluid and of the space-time itself, 
our analysis will give a new insight into the 
formations of naked singularities in non-spherically symmetric 
space-time. 

To decouple physical effects from gauge or coordinate ones, 
we adopt the gauge-invariant
formalism formulated by Gerlach and Sengupta\cite{Gerlach-Sengupta}  
for general spherically symmetric space-times. 
Here we consider only the metric perturbations, 
i.e., linear gravitational waves. 
Using this formalism, we 
obtain a single decoupled partial differential equation 
for a gauge-invariant variable corresponding to the 
odd-parity metric perturbations of the LTB space-time. 
We analyze this equation numerically by use of single null 
coordinates, which was adopted by Goldwirth and Piran\cite{Goldwirth} 
for the numerical study of spherical collapse 
of a massless scalar field. 
Then we shall discuss the stability of the LTB space-time with 
a central naked singularity from the results of these analyses. 
A naked singularity is interpreted to be unstable, 
if perturbations tend to diverge as they approach 
the naked singularity and 
the Cauchy horizon associated with it. 
If such a behavior is found, it means that 
the perturbations will destroy the Cauchy horizon and 
change the causal structure of this space-time. 
Waugh and Lake examined the stability
of the central naked singularity of the LTB 
space-time against perturbations of a massless field
by the use of the high-frequency (eikonal) 
approximation and with the assumption of no back reaction 
of the massless field to the space-time geometry\cite{Waugh-Lake}. 
Their analysis revealed that the formation of the 
central naked singularity 
is stable within the validity of their approximation. 
In contrast to the analysis by Waugh and Lake, however, 
the effect of the finite wavelength of the perturbations and the 
non spherically symmetric dynamics of the space-time itself 
are taken into account up to the linear order in our present analysis. 

This paper is organized as follows. 
In Section \ref{sec:LTB space-time}, we briefly 
describe the LTB space-time. In Section \ref{sec:perturbation}, 
we derive the basic equations for perturbations 
in the LTB space-time and 
then give regularity conditions 
for the perturbations at the symmetric center, 
which are significant for our stability analysis of the central 
naked singularity formation. 
In Section \ref{sec:Riemann tensor}, we present the expressions 
for the perturbations of the Riemann tensor. 
We show the numerical procedure and 
results for the marginally bound LTB space-time 
in Section \ref{sec:numerical results}. 
Section \ref{sec:conclusions} is devoted 
to the summary and discussions. 
We adopt the geometrized units, $c=G=1$. 
The signature of the metric tensor and 
sign convention of the Riemann tensor follow 
Ref.\cite{MTW}. 

\section{LEMA\^{I}TRE-TOLMAN-BONDI SPACE-TIME}
\label{sec:LTB space-time}
The inhomogeneous spherically symmetric dust 
collapse is described by the LTB space-time. 
Using the  synchronous coordinate system, the 
line element of this space-time is expressed in the form  
\begin{equation}
  \label{bgmetric}
  d{\bar s}^{2}= {\bar g}_{\mu\nu}dx^{\mu}dx^{\nu}\equiv
  -dt^2+A^{2}(t,r)dr^{2}
  +R^2(t,r)(d\theta^{2}+\sin^{2}\theta d\phi^{2}).
\end{equation}
The energy-momentum tensor for the dust fluid is 
\begin{equation}
  \label{bgmatter}
  {\bar T}^{\mu\nu} = {\bar \rho}(t,r){\bar u}^{\mu}{\bar u}^{\nu},
\end{equation}
where ${\bar\rho}(t,r)$ is the rest mass density and 
${\bar u}^{\mu}$ is the 4-velocity of the dust fluid. 
In the synchronous coordinate system, the unit normal vector field to 
the spacelike hypersurfaces  is geodesic and  
there is a freedom of which timelike geodesic field 
is adopted as the hypersurface unit normal. 
Using this freedom, we can always set $u^{\mu}=\delta^{\mu}_{0}$ 
since the 4-velocity of the spherically symmetric 
dust fluid is tangent to an irrotational 
timelike geodesic field. 

Then the Einstein equations and the equation of motion for the 
dust fluid reduce to the following simple equations, 
\begin{eqnarray}
  A &=&  \frac{\partial_{r}R}{\sqrt{1+f(r)}}, \label{eq:A} \\
  {\bar \rho}(t,r) &=& \frac{1}{8\pi}
  \frac{1}{R^2 \partial_{r}R}{dF(r)\over dr},\label{eq:einstein} \\
  (\partial_{t}R)^2-\frac{F(r)}{R} &=& f(r),\label{eq:energyeq} 
\end{eqnarray}
where $f(r)$ and $F(r)$ are arbitrary functions of the radial 
coordinate, $r$. From Eq.(\ref{eq:einstein}), $F(r)$ is related 
to the Misner-Sharp mass function\cite{Misner}, $m(r)$, 
of the dust cloud in the manner 
\begin{equation}
  \label{mass}
  m(r) = 4\pi \int_0^{R(t,r)}{\bar \rho}(t,r)R^2dR = 4\pi
  \int_0^r{\bar \rho}(t,r)R^2\partial_{r}R dr =\frac{F(r)}{2}. 
\end{equation}
Hence Eq.(\ref{eq:energyeq}) might be 
regarded as the energy equation per unit mass. 
This means that  
the other arbitrary function, $f(r)$, is recognized as 
the specific energy of the dust fluid. 
The motion of the dust cloud is completely specified 
by the function, $F(r)$, 
(or equivalently, the initial distribution of 
the rest mass density, ${\bar \rho}$) and the specific energy, $f(r)$. 
When we restrict our calculation to the case that the symmetric center, 
$r=0$, is initially regular, the central shell focusing singularity 
is naked if and only if 
$\partial_{r}^{2}{\bar\rho}|_{r=0}<0$ is initially 
satisfied for the marginally bound collapse, $f(r)=0$
\cite{Singh-Joshi,Jhingan}. 
For the collapse that is not marginally bound, there exists a similar
condition as an inequality for a value depending on the functional forms 
of $F(r)$ and $f(r)$\cite{Newman,Singh-Joshi,Jhingan}.
\section{PERTURBATION METHOD}
\label{sec:perturbation}
The perturbation method used in this article is the gauge-invariant
one which has been formulated by Gerlach and Sengupta\cite{Gerlach-Sengupta}
for a general spherically symmetric background space-time. 
First we briefly review their formalism only 
for the so-called odd-parity modes. 
Thereafter we will apply this formalism to the  
LTB space-time in order to derive the basic equations for our 
analysis. 

The perturbed metric tensor is expressed in the form 
\begin{equation}
  \label{metric}
  g_{\mu\nu} = {\bar g}_{\mu\nu}+h_{\mu\nu},
\end{equation}
where ${\bar g}_{\mu\nu}$ is the metric tensor of
the spherically symmetric background space-time 
and $h_{\mu\nu}$ is a perturbation. 
The energy-momentum tensor is written in the form  
\begin{equation}
  \label{ene-mom}
  T_{\mu\nu} = \bar{T}_{\mu\nu} + \delta T_{\mu\nu},
\end{equation} 
where $\bar{T}_{\mu\nu}$ is a background quantity and 
$\delta T_{\mu\nu}$ is a perturbation. 
By virtue of the spherical symmetry of the background space-time, 
${\bar T}_{\mu\nu}$ is expressed in the form 
\begin{eqnarray} 
  \label{GS-matter} 
  \bar{T}_{\mu\nu}dx^{\mu}dx^{\nu} &=& 
  \bar{T}_{ab}dx^adx^b+\frac{1}{2}\bar{T}_B^{~B}R^2(t,r)d\Omega^2, 
\end{eqnarray} 
where the sub- and superscripts, $a,b,\ldots$ represent $t$ and $r$  
while $A,B,\ldots$ represent $\theta$ and $\phi$. 
The odd-parity perturbations of  
$h_{\mu\nu}$ and $\delta T_{\mu\nu}$ are expressed 
in the form 
\begin{eqnarray} 
  \label{GS-pmetric} 
  h_{\mu\nu} &=& \left(\begin{array}{ccc} 
                        0 & 0 & h_0(t,r)\Phi^m_{l~B} \\ 
                          & 0 & h_1(t,r)\Phi^m_{l~B} \\ 
                     \mbox{Sym} &   & h_2(t,r)\chi^m_{l~AB} 
                      \end{array}  \right), 
\end{eqnarray} 
\begin{eqnarray} 
  \label{GS-pmatter} 
  \delta T_{\mu\nu} &=& \left(\begin{array}{ccc} 
                        0 & 0 & t_0(t,r)\Phi^m_{l~B} \\ 
                          & 0 & t_1(t,r)\Phi^m_{l~B} \\ 
                     \mbox{Sym} &   & t_2(t,r)\chi^m_{l~AB} 
                      \end{array}  \right) , 
\end{eqnarray}  
where $\Phi^m_{l~B}$ and $\chi^m_{l~AB}$ are odd-parity vector and 
tensor harmonics associated with the spherical symmetry of 
the background space-time\cite{Regge}. 

We introduce gauge-invariant 
variables defined by Gerlach and Sengupta. The metric 
variables are given by 
\begin{equation} 
  \label{ka} 
  k_a = h_a-\frac{1}{2}R^2\partial_a\left(\frac{h_2}{R^2}\right).
\end{equation} 
The matter variables are given by the following combinations, 
\begin{eqnarray}
  \label{la}
  L_a &=& t_a-\frac{1}{2}T_B^{~B}h_a ,\\
  \label{l}
  L &=& t_2-\frac{1}{2}T_B^{~B}h_2 .
\end{eqnarray}
Then the Einstein equations lead to the equations  
for the metric variables as 
\begin{eqnarray}
  \label{GS-9a}
  k^a_{~|a} &=& 16 \pi L~~~~~~~(l\geq 2),
\end{eqnarray}
\begin{eqnarray}
  \label{GS-9b}
  \left(R^4W^{ab}\right)_{|b}+\left(l-1\right)\left(l+2\right)k^a &=&
  16 \pi R^2L^a~~~~~~~(l\geq 1),
\end{eqnarray}
where $W_{ab}$ is defined as
\begin{equation}
  \label{Wab}
  W_{ab} \equiv
  \left(\frac{h_b}{R^2}\right)_{|a}-\left(\frac{h_a}{R^2}\right)_{|b}
         =
  \left(\frac{k_b}{R^2}\right)_{|a}-\left(\frac{k_a}{R^2}\right)_{|b},
\end{equation}
and the vertical bar refers to the covariant derivative 
within the 2-dimensional sub-space-time $(t,r)$. From the equation 
of motion for the matter, we get 
\begin{equation}
  \label{GS-14}
  \left(R^2L^a\right)_{|a} =
  \left(l-1\right)\left(l+2\right)L~~~~~~~(l\geq 1).
\end{equation} 

Now we apply the above formalism to the case of the 
background LTB space-time. From Eqs.(\ref{bgmatter}) and (\ref{GS-pmatter}), 
we find that there is no density 
perturbation and that only the perturbation of 4-velocity, 
$\delta u_{\mu}$, exists; 
\begin{equation}
  \delta u_{\mu} = (0,0,U(t,r)\Phi^m_{l~B}).
\end{equation}
Therefore the odd-parity gauge-invariant
matter variables become 
\begin{eqnarray}
  \label{La-L}
      L_0 = \bar{\rho} U~~~~~~{\rm and}~~~~~~ L_1 = L = 0.
\end{eqnarray}
 From Eqs.(\ref{GS-9a}) and (\ref{GS-9b}), we obtain 
the equations of motion for the metric variables, 
 \begin{eqnarray}
  \label{wave1}
  \partial_t\left(Ak_0\right)
  -\partial_r\left(\frac{k_1}{A}\right) &=& 0, \\
  \label{wave2}
  \partial_r\left(R^4\psi_s\right)+A\left(l-1\right) 
\left(l+2\right)k_0 &=& 16\pi AR^2L_0, \\
  \label{wave3}
  \partial_t\left(R^4\psi_s\right)+\frac{1}{A}
\left(l-1\right)\left(l+2\right)k_1 &=& 0,  
\end{eqnarray}
where we have introduced another gauge-invariant variable, $\psi_s$,
defined as 
\begin{equation}
  \label{psis}
  \psi_s \equiv {1\over A}\left[\partial_t
  \left(\frac{k_1}{R^2}\right)-\partial_r
  \left(\frac{k_0}{R^2}\right)\right].
\end{equation}

Eq.(\ref{GS-14}) becomes
\begin{equation}
  \label{T-Bconserve}
  \partial_t\left(AR^2L_0\right) = 0.
\end{equation}
This equation is easily integrated and we obtain
\begin{equation}
  \label{kakuundouryou}
  AR^2L_0 = {dJ(r)\over dr},
\end{equation}
where $J(r)$ is an arbitrary function depending only on 
$r$. From Eqs.(\ref{wave2}), (\ref{wave3}), 
and (\ref{kakuundouryou}), 
we obtain a single decoupled wave equation in the form 
\begin{equation}
  \label{wave4}
  \partial_{t}\left(\frac{A}{R^2}\partial_{t}
\left(R^4\psi_s
\right)\right)-\partial_{r}
\left(\frac{1}{AR^2}\partial_{r}
\left(R^4\psi_s\right)\right)+
\left(l-1\right)
\left(l+2\right)A\psi_s 
= -16\pi\partial_{r}\left(\frac{1}{AR^2}{dJ\over dr}\right).
\end{equation}

The variable, $\psi_{s}$, differs from the
gauge-invariant variable used in Ref.\cite{Seidel} by a 
factor of $1/R^2$.
The reason why we adopt $\psi_s$ as a gauge-invariant variable 
instead is due to regularity conditions at  
$r=0$. Further, as will be shown later, $\psi_s$ 
is closely connected with the curvature tensor near the center. 

Let us consider the regularity conditions for the background 
metric functions and gauge-invariant perturbations at $r=0$. 
Hereafter we restrict ourselves to the axisymmetric case, 
i.e., $m=0$. Note that this restriction does not lose generality 
of our analysis. Further we consider only the case in which 
the space-time is regular before the occurrence of the singularity. 
This means that, before the naked singularity 
formation, the metric functions, $R(t,r)$ and $A(t,r)$, behave 
near the center in the manner 
\begin{eqnarray}
  \label{r-0-R}
  R &\longrightarrow& R_c(t)r+O(r^3), \\
  \label{r-0-A}
  A &\longrightarrow& R_{c}(t)+O(r^{2}).
\end{eqnarray}
To investigate the regularity conditions of the gauge-invariant 
variables, $k_a$ and $L_0$, 
we follow Bardeen and Piran\cite{Bardeen}. The results are given by 
\begin{eqnarray}
  \label{r-0-L}
  L_0 &\longrightarrow& L_c(t)r^{l+1}+O(r^{l+3}), \\
  \label{r-0-k0}
  k_0 &\longrightarrow& k_{0c}(t)r^{l+1}+O(r^{l+3}), \\
  \label{r-0-k1}
  k_1 &\longrightarrow& k_{1c}(t)r^{l+2}+O(r^{l+4}). 
\end{eqnarray}
 From Eqs.(\ref{psis}), (\ref{r-0-R}), (\ref{r-0-A}),(\ref{r-0-k0}) and 
(\ref{r-0-k1}), we find that $\psi_s$ behaves near the center as 
\begin{eqnarray}
  \label{r-0-psis}
  \psi_{s} &\longrightarrow& 
 \psi_{sc}(t)r^{l-2}+O(r^l)
 ~~~~~~~~~~~~\mbox{for $l\geq 2$}, \\
  \label{r-0-psis1}
  \psi_{s} &\longrightarrow& \psi_{sc(t)}r+O(r^3)
  ~~~~~~~~~~~~~~~~\mbox{for $l=1$}.
\end{eqnarray}
In the case of $l\geq2$, the coefficient, $\psi_{sc}(t)$, 
is related to $R_{c}(t)$ and $k_{0c}(t)$ in the manner 
\begin{equation} 
  \label{relation-psis}
\psi_{sc}(t)=-(l-1)\frac{k_{0c}(t)}{R_{c}^{3} (t)}.
\end{equation}
 From the above equations, we note that only the quadrupole mode, 
$l=2$, of $\psi_{s}$ does not vanish at the center. 

\section{PERTURBATION OF RIEMANN TENSOR}
\label{sec:Riemann tensor}
In this section, we consider the perturbation of the Riemann tensor,
$R_{\mu\nu\sigma}^{~~~~\lambda}$, of the 
LTB space-time to investigate the relation between the singularity
formation and the perturbations. The Riemann tensor is 
decomposed into the Ricci tensor, $R_{\mu\nu}$, and the Weyl 
tensor, 
\begin{equation}
  \label{Weyl}
  C_{\mu\nu\sigma\lambda}=R_{\mu\nu\sigma\lambda}
+\{g_{\mu[\lambda}R_{\sigma]\nu}
+g_{\nu[\sigma}R_{\lambda]\mu}\}
+\frac{1}{3}Rg_{\mu[\sigma}g_{\lambda]\nu}.
\end{equation}
We shall give them in the form of the components 
of the following tetrad basis, 
\begin{eqnarray}
  \label{tetrad}
  e^{\mu}_{(t)} &=& \left(1,0,0,-\frac{h_0 P_{l,\theta}}{R^2 \sin
  \theta}\right), \\
  e^{\mu}_{(r)} &=& \left(0,\frac{1}{A},0,-\frac{h_1 P_{l,\theta}}{AR^2  \sin
  \theta}\right), \\
  e^{\mu}_{(\theta)} &=&
  \left(0,0,\frac{1}{R},-\frac{h_2}{2R^3\sin^2\theta} \left(\sin\theta
  P_{l,\theta,\theta}-\cos\theta P_{l,\theta}\right)\right), \\
  e^{\mu}_{(\phi)} &=& \left(0,0,0,\frac{1}{R\sin \theta}\right),
\end{eqnarray}
where $P_{l}(\cos \theta)$ is the Legendre polynomial 
and the comma followed by $\theta$ denotes a  
derivative with respect to $\theta$.
The Weyl tensor is then decomposed into the so-called 
electric part, $E_{\alpha\beta}$, and 
magnetic part, $B_{\alpha\beta}$, which are defined as
\begin{eqnarray}
  \label{electric-part}
  E_{\alpha \beta} &\equiv& C_{\alpha \mu \beta \nu}
  e_{(t)}^{\mu}e^{\nu}_{(t)},\\
  \label{mag-part}
  B_{\alpha \beta} &\equiv& \frac{1}{2}
  {\epsilon_{\alpha\sigma}}^{\mu
  \nu}C_{\mu \nu \beta \lambda}e^{\sigma}_{(t)}e^{\lambda}_{(t)},
\end{eqnarray}
where $\epsilon_{\mu\nu\alpha\beta}$ is the 4-dimensional skew tensor. In
the background LTB space-time, the Ricci tensor has a non-zero value
in the region of non-vanishing rest mass density, ${\bar\rho} \neq 0$, 
through the Einstein equations and also
the electric part has a non-zero value. 
On the other hand, the magnetic part 
is identically equal to zero in the background LTB 
space-time. 
However, when axisymmetric odd-parity metric perturbations exist, 
the Riemann tensor is perturbed and 
the magnetic part may also have 
a non-vanishing value. 

The perturbation of the Ricci tensor is expressed by the matter
perturbation through the Einstein equations as 
\begin{eqnarray}
  \label{p-ricci}
  \delta(R_{(t)(\phi)}) = \frac{8\pi}{R}L_0P_{l,\theta} = 
\frac{8\pi}{AR^3}\frac{dJ}{dr}P_{l,\theta},
\end{eqnarray}
and the other components vanish, where we have used 
Eq.(\ref{kakuundouryou}) 
in the last equality.
 The perturbations of the tetrad components of the
electric part are given in the form 
\begin{eqnarray}
  \label{p-ele-part}
  \delta(E_{(r)(\phi)}) &=&
  {1\over2}\left[\frac{1}{AR^3}(l-1)(l+2)k_1
  +R(\partial_{t}R)\psi_s\right]
\sin\theta P_{l,\theta}, \\
  \delta(E_{(\theta)(\phi)}) &=&
  \frac{1}{2AR^2}\left[\partial_{t}\left(\frac{k_1}{A}\right)
  -(\partial_{t}A)k_0\right]
\left(\sin \theta P_{l,\theta,\theta}-\cos\theta P_{l,\theta}\right),
\end{eqnarray}
and the other components vanish. The perturbations of the tetrad
components of the magnetic part are obtained 
in the form
\begin{eqnarray}
  \label{p-mag-part}
\delta(B_{(r)(r)}) &=& \frac{1}{2}l(l+1)\psi_sP_l, \\
\delta(B_{(r)(\theta)}) &=&
  \frac{1}{4AR^3}\left[R^2\partial_{r}\left(R^{2}\psi_s\right)
  -A(l-1)(l+2)k_0\right]
  P_{l,\theta}, \\
\delta(B_{(\theta)(\theta)}) &=&
  -\frac{1}{AR^2}\left[R\partial_{r}\left(\frac{k_0}{R}\right)
  +\left(\frac{\partial_{t}R}
  {R}-\frac{\partial_{t}A}{A}\right)k_1+\frac{1}{2}AR^{2}\psi_s\right]
  P_{l,\theta,\theta} 
\nonumber\\
  &  &-\frac{1}{2AR^2}\left[R\partial_{r}\left(\frac{k_0}{R}\right)
  +\left(\frac{\partial_{t}R}
  {R}-\frac{\partial_{t}A}{A}\right)k_1+AR^{2}\psi_s\right]l(l+1)P_l, \\
\delta(B_{(\phi)(\phi)}) &=&
  -\frac{1}{AR^2}\left[R\partial_{r}\left(\frac{k_0}{R}\right)
  +\left(\frac{\partial_{t}R}
  {R}-\frac{\partial_{t}A}{A}\right)k_1+\frac{1}{2}AR^{2}\psi_s\right]
  \cot\theta P_{l,\theta} \nonumber \\
 \label{p-mag-pp}
  &  &-\frac{1}{2AR^2}\left[R\partial_{r}\left(\frac{k_0}{R}\right)
  +\left(\frac{\partial_{t}R}
  {R}-\frac{\partial_{t}A}{A}\right)k_1+AR^{2}\psi_s\right]l(l+1)P_l,
\end{eqnarray}
and the other components vanish.

Now we will investigate the behavior of the Ricci and Weyl tensors near
the center where the naked singularity appears. From the regularity
conditions (\ref{r-0-R})-(\ref{relation-psis}), we can see that the 
perturbations of the Ricci and Weyl
tensors obtained in the above behave 
near the center in the manner 
\begin{eqnarray}
  \label{r-0-ricci}
  \delta(R_{(t)(\phi)}) &\longrightarrow& \frac{8\pi}{R_c}L_c P_{l,\theta}
  r^l,
\end{eqnarray}
for the Ricci tensor, and  
\begin{eqnarray}
  \label{r-0-ela-part}
  \delta(E_{(r)(\phi)}) &\longrightarrow&
  \frac{1}{2R^4_c}(l-1)\left[(l+2)k_{1c}-R_c{dR_c\over dt}k_{0c}\right]
  P_{l,\theta}r^{l-1}, \\
  \delta(E_{(\theta)(\phi)}) &\longrightarrow&
  \frac{1}{2R^4_c}\left[(l+2)k_{1c}-R_c{dR_c\over dt}k_{0c}\right]
(P_{l,\theta,\theta}-\cot \theta P_{l,\theta})r^{l-1}, \\
  \label{r-0mag-part}
  \delta(B_{(r)(r)}) &\longrightarrow&
  -\frac{1}{2R^3_c}(l-1)l(l+1)k_{0c}P_lr^{l-2}, \\
  \delta(B_{(r)(\theta)}) &\longrightarrow&
  -\frac{1}{2R^3_c}(l-1)(l+1)k_{0c}P_{l,\theta}r^{l-2}, \\
  \delta(B_{(\theta)(\theta)}) &\longrightarrow&
  -\frac{1}{2R^3_c}(l+1)k_{0c}(P_{l,\theta,\theta}+lP_l)r^{l-2}, \\
  \delta(B_{(\phi)(\phi)}) &\longrightarrow&
  -\frac{1}{2R^3_c}(l+1)k_{0c}(\cot\theta P_{l,\theta}+lP_l)r^{l-2},
\end{eqnarray}
for the Weyl tensor of $l\geq2$. 
For the $l=1$ mode, we find 
\begin{eqnarray}
  \label{r-0-l-1}
  \delta(E_{(r)(\phi)}) &\longrightarrow&
  \frac{1}{2}{dR_c \over dt}\psi_{sc}r^2\sin\theta, \\
  \delta(B_{(r)(r)}) &\longrightarrow& 
  -\psi_{sc}r\cos\theta,\\ 
  \delta(B_{(r)(\theta)}) &\longrightarrow&
  \frac{1}{4}\psi_{sc}r\sin\theta, \\
  \delta(B_{(\theta)(\theta)}) &\longrightarrow&
  \frac{1}{2}\psi_{sc}r\cos\theta, \\
  \delta(B_{(\phi)(\phi)}) &=&   \delta(B_{(\theta)(\theta)}). 
\end{eqnarray}

 From the above equations, we see that 
the perturbations of the tetrad components of the 
Ricci and Weyl tensors, except for 
the quadrupole mode, $l=2$, 
of the magnetic part, $B_{\alpha\beta}$, identically vanish 
at the center. This means that the central naked singularity 
formation is affected only by the quadrupole mode up to linear order.
Therefore, hereafter we shall consider the quadrupole mode only. 
On the other hand, since the solution for the dipole mode, $l=1$, 
is obtained analytically, we present it in Appendix \ref{sec:Dipole},
although the dipole mode vanishes at the center from the regularity 
condition and does not influence the formation of the central 
naked singularity up to linear order. 

\section{NUMERICAL RESULTS AND DISCUSSIONS}
\label{sec:numerical results}
Here we will perform numerical integration of Eq.(\ref{wave4}) 
for the quadrupole mode, $l=2$. As mentioned above, the gauge-invariant 
matter perturbation variable, $L_0$, and the perturbation of the Ricci
tensor vanish at the regular center. 
Here we restrict our investigation to the case 
of no matter perturbations, namely the right
hand side of Eq.(\ref{wave4}) vanishes. The non-vanishing matter
perturbation case should be investigated in future.
Further, for the simplicity of calculations, we 
consider only the marginally bound case, $f(r)=0$, as 
background space-time. 

By virtue of $f(r)=0$, 
we can easily integrate Eq.(\ref{eq:energyeq}) and obtain 
\begin{equation}
  \label{f=0}
  R(t,r) = \left(\frac{9F}{4}\right)^{1/3}[t_0(r)-t]^{2/3},
\end{equation}
where $t_0(r)$ is an arbitrary function of 
$r$. Using the freedom
for the scaling of $r$, we choose $R(0,r)=r$. This scaling of $r$ 
corresponds to the following choice of $t_{0}(r)$, 
\begin{equation}
  \label{t0}
  t_0(r) = \frac{2}{3\sqrt{F}}r^{3/2}.
\end{equation}
Here note that, from Eq.(\ref{eq:A}), the background 
metric variable, $A$, is equal to $\partial_{r}R$.  

Then, the wave equation (\ref{wave4}) becomes as follows,
\begin{eqnarray}
  \label{wave-eq}
  \frac{\partial^2 \psi_s}{\partial t^2}-\frac{1}{(\partial_{r}R)^2}
 \frac{\partial^2 \psi_ s}{\partial  r^2}
  &=& \frac{1}{(\partial_{r}R)^{2}}\left(6\frac{\partial_{r}R}{R}
  -\frac{\partial_{r}^{2}R}{\partial_{r}R}\right)\frac{\partial
  \psi_s}{\partial r}-\left(6\frac{\partial_{t}R}{R}
  +\frac{\partial_{t}\partial_{r}R}{\partial_{r}R}\right)
  \frac{\partial
  \psi_s}{\partial t} \nonumber \\ 
& &-4\left[\left(\frac{\partial_{t}\partial_{r}R}{\partial_{r}R}\right)
  \frac{\partial_{t}R}{R}
  +\left(\frac{\partial_{t}R}{R}\right)^2+
  \frac{\partial_{t}^{2}R}{R}\right]\psi_s .
\end{eqnarray}
We solve this partially differential equation numerically. In the 
rest of this section, we explain the details of the background space-time 
considered here, numerical methods and boundary
conditions. Further we show the numerical results. 
\subsection{Background Density Profile}
\label{sec:model}
The background space-time and the motion of 
the marginally bound dust cloud are completely  
determined by the initial rest mass density 
profile, ${\bar \rho}(0,r)$. Further as already mentioned, 
it should satisfy the 
condition $\partial_{r}^{2}{\bar\rho}|_{r=0}<0$ 
in order that the central naked singularity is formed.  
We, therefore, adopt the following initial rest mass 
density profile so that 
the central naked singularity appears; 
\begin{equation}
  \label{density}
  {\bar\rho}(0,r) = \left\{ 
      \begin{array}{ccc}
        \rho_0[1-2(r/r_b)^2+(r/r_b)^4] &{\mbox{for}} & 0\leq r\leq r_b\\
        0 &{\mbox{for}} & r>r_b,
      \end{array}
    \right.
\end{equation}
where $\rho_0$ is a positive constant and $r_b$ denotes 
the radial coordinate at the surface of the dust cloud. 
The total (gravitational) mass of the dust cloud is 
\begin{equation}
  \label{totmass}
  M=m(r_b)=\frac{32\pi}{105}\rho_0 r_b^3.
\end{equation}
The time of the central naked singularity formation is 
\begin{equation}
  \label{sing-time}
  t=t_0(0) = \frac{1}{\sqrt{6\pi\rho_0}} .
\end{equation}
Whether the naked singularity is global or local is
determined by a non-dimensional constant $\rho_0r_b^2$. It is
known that the singularity is globally naked for sufficiently 
small $\rho_0r_b^2$\cite{Christodoulou,Singh-Joshi}. 
However, the critical value of $\rho_0r_b^2$ can not be obtained
explicitly. Hence, after $\rho_{0}r_{b}^{2}$ is given, 
we have to investigate 
whether the central naked singularity is global or local, 
by numerically solving the future directed 
null ray from the central naked singularity. 
Here we consider two cases. 
One is that of $\rho_{0}r_{b}^{2}
=3\times10^{-2}$, which corresponds to a globally naked 
singularity, and the other is that of $\rho_{0}r_{b}^{2}=3\times10^{-1}$, 
which corresponds to a locally naked one. 

In the globally naked case, the initial radius of the dust cloud 
and the time of the central naked singularity formation are given by 
\begin{eqnarray}
\label{mrb}
 {R(0,r_b)\over M}
&=& \frac{r_b}{M} = \frac{105}{32\pi\rho_0 r_b^2} \cong 34.8, \\
 \frac{t_0(0)}{M} 
&=& \frac{105}{32\sqrt{6}\pi^{3/2}(\rho_0 r_b^2)^{3/2}} \cong 46.3.
\end{eqnarray}
On the other hand, in the locally naked case, they are given by 
\begin{eqnarray}
\label{mrbl}
  {R(0,r_b)\over M}
&=& \frac{r_b}{M} \cong 3.48, \\
  \frac{t_0(0)}{M} &\cong& 1.46.
\end{eqnarray}

\subsection{Numerical Procedure}
\label{sec:method}
Next, we describe the procedure of our numerical calculation. 
We have a disadvantage when we use the $(t,r)$ coordinate 
system, because
of the restriction on the region in which we can numerically 
construct the solution of the wave equation, (\ref{wave4}). 
Therefore, instead of the $(t,r)$ coordinate system, we 
introduce a single-null coordinate system, $(u,r')$,
where $u$ is an out-going null coordinate and chosen so that it
agrees with $t$ at the symmetric center and we choose $r'=r$.
We perform the numerical integration along two characteristic
directions. The transformation matrix is formally expressed in the form 
\begin{eqnarray}
  \label{dr'dlambda}
    dr' &=& dr,  \\
  \label{u-matrix}
    du  &=& \left(\partial_{t}u\right)_rdt
                +\left(\partial_{r} u\right)_tdr.
\end{eqnarray}
Because $u$ is the out-going null coordinate, the following relation holds,
\begin{equation}
  \label{indl}
  \frac{\left(\partial_{t} u\right)_r}
  {\left(\partial_{r} u\right)_t}
 = -\frac{1}{\partial_{r}R}.
\end{equation}
Using these relations, we obtain the line element of the 2-dimensional 
sub-space-time, $(t,r)$, in the following new form 
\begin{equation}
  \label{metric-affine}
  ds^2_{(2)} = -\alpha^2du^2 - 2\alpha (\partial_{r}R) du dr', 
\end{equation}
where we have introduced 
\begin{equation} 
  \alpha\equiv {1\over (\partial_{t} u)_r}.
\end{equation}
Furthermore, from Eqs.(\ref{dr'dlambda}) and (\ref{u-matrix}), 
we obtain 
\begin{equation}
  \label{r'a}
  \partial_{r'}= -\frac{(\partial_r u)_t}{(\partial_t
    u)_r}\partial_{t}+\partial_{r}=
    (\partial_{r}R)\partial_{t}+\partial_{r},
\end{equation}
where we have used Eq.(\ref{indl}) in the last equality.
The above equation describes that $\partial_{r'}$ 
is parallel to the future directed out-going null direction. 

By using this new coordinate system, $(u,r')$, Eq.(\ref{wave-eq}) 
is expressed in the form 
\begin{eqnarray}
  \label{dphis/dlambda}
  \frac{d\phi_s}{du} &=&
  -\frac{\alpha}{R}\left[3\partial_{r}R
  +{1\over2}R(\partial_{t}R)\partial_{t}\partial_{r}R
  -(\partial_{t}R)^{2}\partial_{r}R
  +{1\over2}R(\partial_{r}R)\partial_{t}^{2}R\right]\psi_s \nonumber \\
    & &
  -\frac{\alpha}{2}\left[\frac{\partial_{r}^{2}R}{(\partial_{r}R)^2}
  -\frac{2}{R}\left(1-\partial_{t}R\right)
\right]\phi_s , \\
  \label{delpsi/delrprime}
  \partial_{r'}\psi_{s}&=& \frac{1}{R}\phi_s
  -3\frac{\partial_{r}R}{R}
\left(1+\partial_{t}R\right)\psi_s,
\end{eqnarray}
where the ordinary derivative in the left hand side of
Eq.(\ref{dphis/dlambda}) is given by
\begin{equation}
  \frac{d}{du} = \partial_u +
  \frac{dr'}{du}\partial_{r'}
   = \partial_u-\frac{\alpha}{2\partial_r R}\partial_{r'},
\end{equation}
and we have introduced a new variable, $\phi_s$, given by 
Eq.(\ref{delpsi/delrprime}).

The procedure of the numerical integration is as
follows. At the first step, we prepare initial data 
corresponding to imploding waves for 
$\phi_s$ at each grid point on the initial null hypersurface 
labeled by $u=u_0=$constant.  
Then, using this $\phi_s$, $\psi_s$ is obtained at each
grid point on $u=u_0$ by the integration of 
Eq.(\ref{delpsi/delrprime}). 
At the next step, in order to obtain $\phi_s$ at each grid point 
on $u=u_0+\Delta u$, we integrate Eq.(\ref{dphis/dlambda}) 
by using values of $\phi_s$ and $\psi_s$ on $u=u_0$. 
Then, $\psi_s$ on $u=u_0+\Delta u$ is
obtained from Eq.(\ref{delpsi/delrprime}) by 
using $\phi_s$ on $u=u_0+\Delta u$. 
We repeat this procedure and obtain a solution outside the 
Cauchy horizon associated with the central naked singularity. 

In the above procedure, we should impose a boundary condition on 
$\psi_s$ at the center to perform the numerical integration 
of Eq.(\ref{delpsi/delrprime}). From Eqs.(\ref{r-0-psis}) 
and (\ref{relation-psis}), 
in the case of $l=2$ mode, $\psi_s$ behaves in the manner 
\begin{equation}
  \label{r-0-psis2}
  \psi_{s} \longrightarrow -\frac{k_{0c}(t)}{R_{c}^{3} (t)}+O(r^2) 
~~~~~{\rm for}~~~~r\longrightarrow 0.
\end{equation} 
Hence, we have to numerically make $\psi_s$ near 
the center so that Eq.(\ref{r-0-psis2}) 
is guaranteed on a surface of $t=$constant, and this leads 
to the boundary condition for $\psi_{s}$ at the center. 

Here we comment on our numerical code. 
We compared the numerical results
for the Minkowski space-time with the analytical solutions which will be
described in the next subsection. In this case, the result produced by
our code agreed with the analytical solution very closely. Another
check we performed was to compare the numerical results for several mesh 
sizes with each other. This test confirmed that 
our numerical results were almost independent of the mesh size. 

\subsection{Initial Conditions and Numerical Results}
\label{sec:results}
The initial conditions which we consider are a  
Gaussian-shaped wave packet with respect to the coordinate, $r'$,
\begin{equation}
  \label{initial}
  \psi_s|_{u=u_{0}} = \psi_s^i \exp\left[-\frac{(r'-r'_c)^2}{2\sigma^2}\right],
\end{equation}
where $\psi_{s}^{i}$, $\sigma$, and $r_{c}'$ are constants 
and characterize the amplitude, width and 
initial position of the initial wave packet, respectively.  
The initial null hypersurface, $u=u_{0}$, is chosen so that 
it includes a world point $(t,r)=(0,0)$, except for the
analysis of the scattered waves which will be discussed in this section. 

\begin{figure}[htbp]
  \begin{center}
    \leavevmode
    \epsfysize=300pt\epsfbox{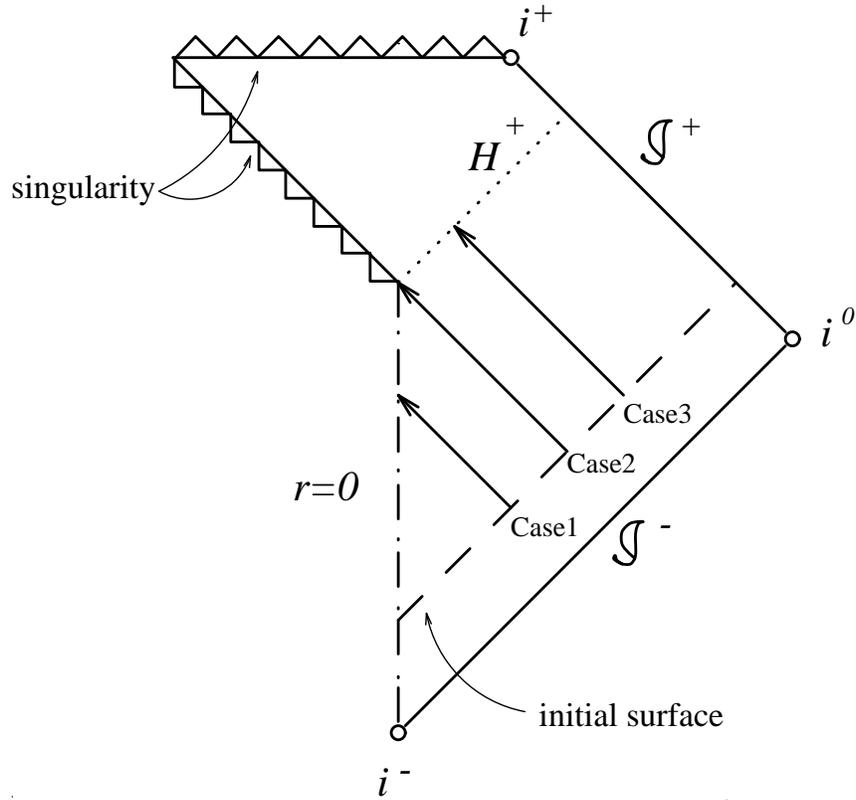}
    \caption{Conformal diagram of the LTB space-time with a globally naked
    singularity. $i^+(i^-)$ denotes future (past) timelike infinity
    respectively, while $i^0$ denotes spacelike
    infinity. $\scri^+(\scri^-)$ denotes future (past) null infinity 
    respectively. The dotted line 
    $H^+$ indicates a future Cauchy horizon associated 
    with the central naked singularity. 
    The broken line is a null
    hypersurface on which we put initial wave packets. 
    The initial positions of the wave packets are classified into 
    Cases 1-3. For the locally naked singularity case, 
    the Cases 1-3 are defined in the same manner as 
    the globally naked case. }
    \label{fig:cases}
  \end{center}
\end{figure}

\begin{figure}[htbp]
  \begin{center}
    \leavevmode
   \begin{tabular}{c}
 \subfigure[Case1,2,3]{\epsfysize=220pt\epsfbox{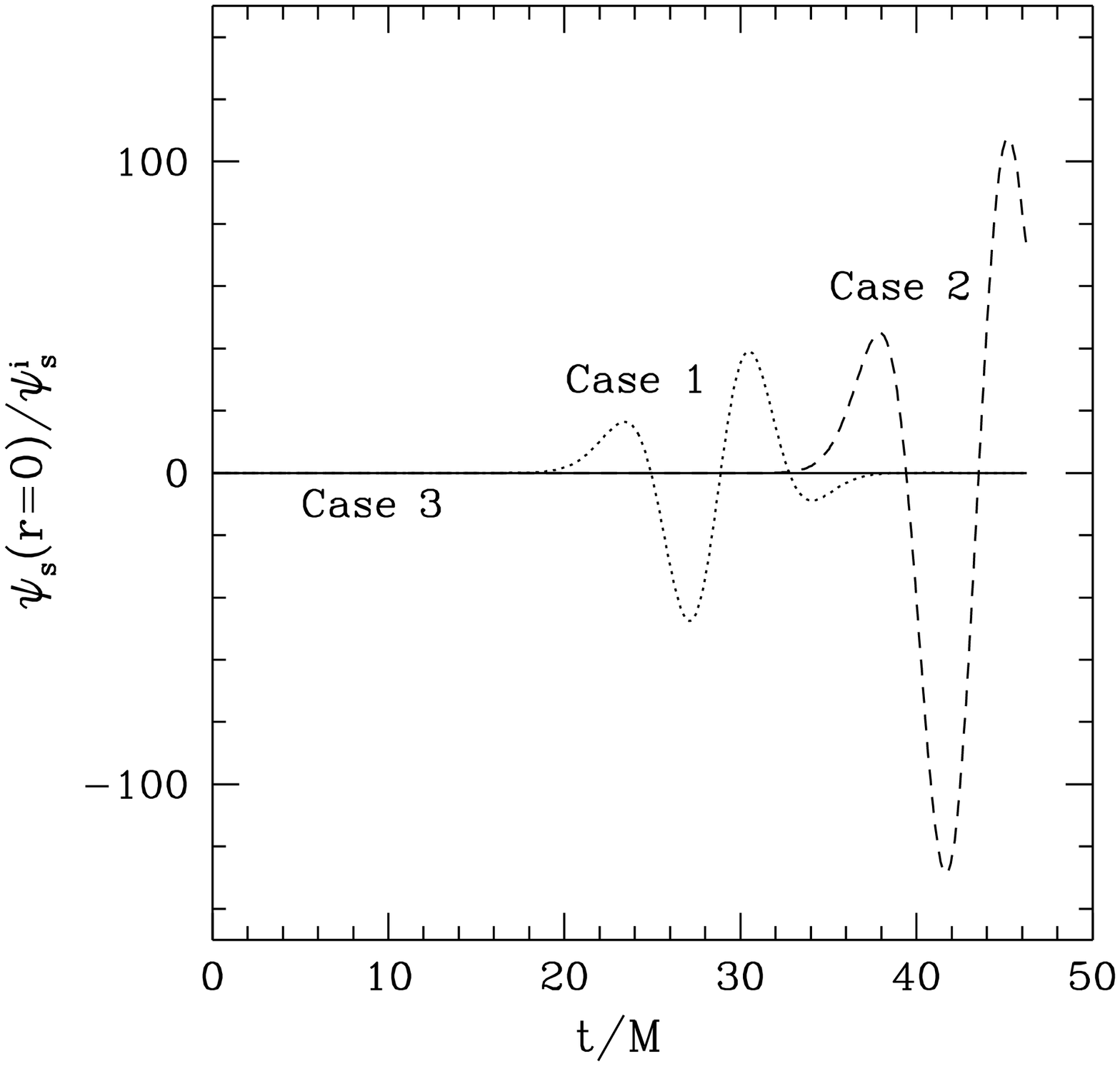}}
 \subfigure[Case2]{\epsfysize=220pt\epsfbox{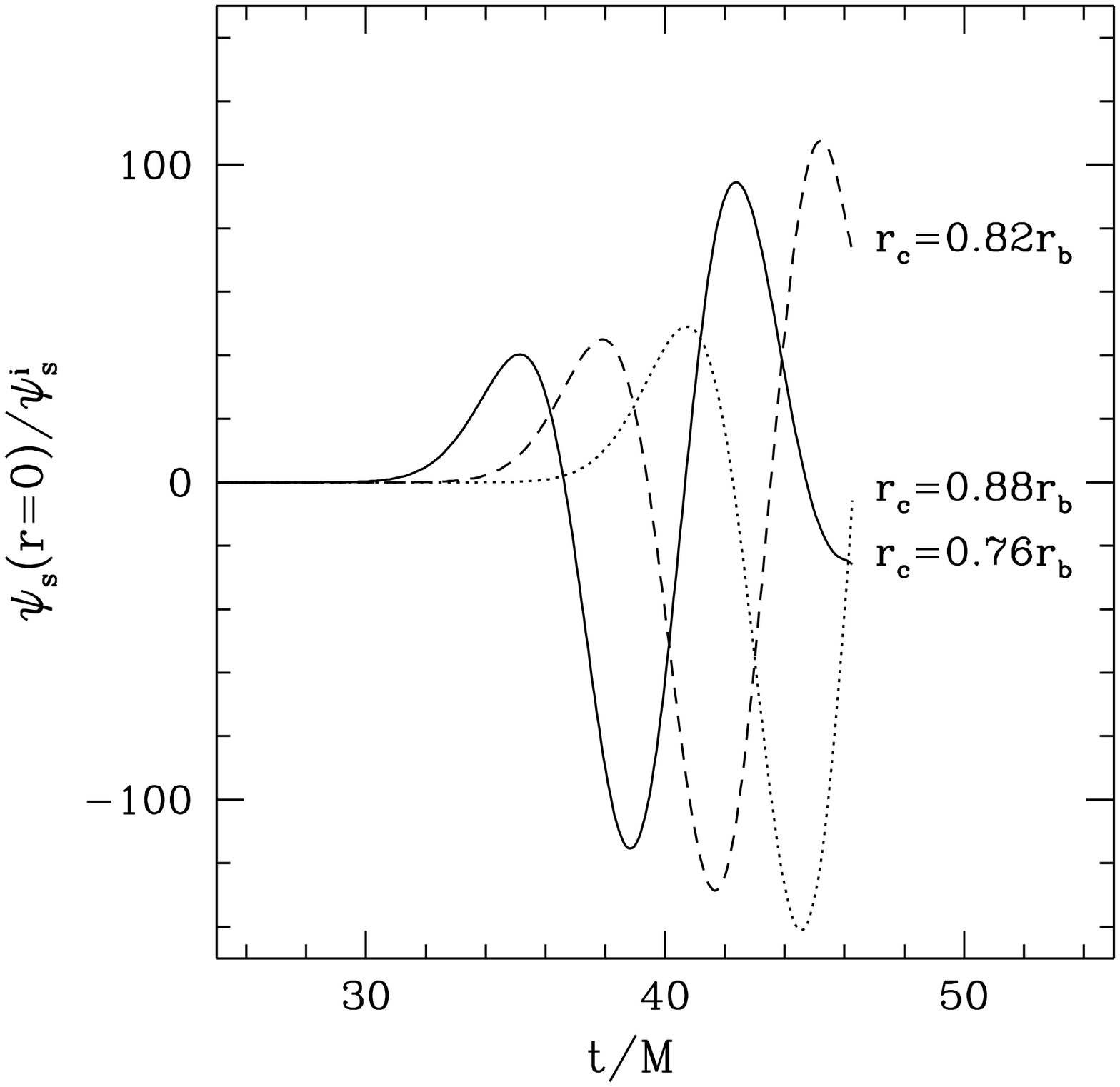}}  
 \end{tabular}
    \caption{Plots of the gauge-invariant variable, $\psi_s$, at the
 center, $r=0$, with an initial width $\sigma = 0.05r_b$ for the Cases
 1-3 for globally naked cases. In
 (a), the dotted line denotes Case 1 $(r'_c=0.5r_b)$, 
 the broken line shows
 Case 2 $(r'_c=0.82r_b)$, and the solid line  
 denotes Case 3 $(r'_c=1.2r_b)$. 
 In (b), the
 results of Case 2 are shown in more detail. 
 The broken line in (b) is the same as the
 broken line in (a). The solid and dotted lines show the cases that
 the wave packets were put on the initial surface at $r'_c=0.76r_b$ and at
 $r'_c=0.88r_b$, respectively.} 
    \label{fig:center1}
  \end{center}
\end{figure}

\begin{figure}[htbp]
  \begin{center}
    \leavevmode
    \epsfysize=220pt\epsfbox{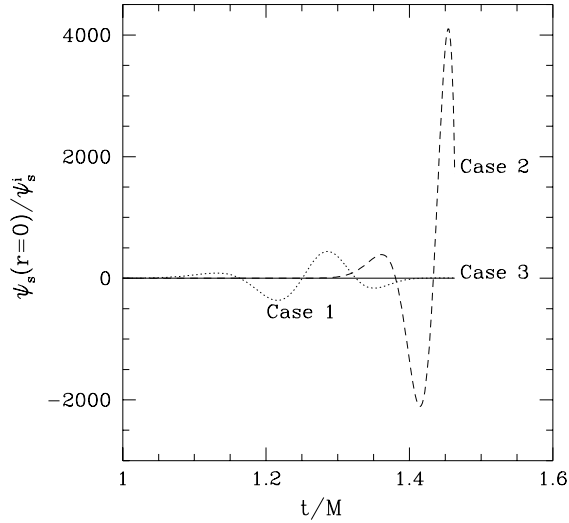}  
    \caption{Plots of the gauge-invariant variable, $\psi_s$, at the
 center, $r=0$, with an initial width $\sigma = 0.02r_b$ for Cases
 1-3 for locally naked cases. The dotted, broken, and solid lines denote
 Case 1 $(r'_c=0.28r_b)$, Case 2 $(r'_c=0.38r_b)$, and Case 3
 $(r'_c=0.58r_b)$, respectively.}
    \label{fig:center2}
  \end{center}
\end{figure}

We investigate models with three different initial positions 
of the wave packet, i.e., $r_{c}'$ in Eq.(\ref{initial}), 
on the initial null hypersurface. 
In Case 1, the wave packet reaches the center of the dust cloud 
before the formation of the central naked singularity. 
In Case 2, a significant portion of the wave packet 
hits the central naked singularity. 
In Case 3, the packet does not hit 
the central naked singularity but reaches the Cauchy horizon 
associated with it.  
Fig.\ref{fig:cases} shows these situations schematically. 
In each case, the value of $\psi_s$ at the center 
is plotted as a function of the coordinate time, $t$, 
in Fig.\ref{fig:center1} for the globally naked case 
and in Fig.\ref{fig:center2} for the locally naked case. 
Note that  it is impossible 
to perform the numerical calculation in the causal future
of the central naked singularity. 
Therefore we plot $\psi_s$ at the center only before the occurrence
of the central naked singularity. 
Although such a difficulty exists, we 
find that violent 
growth of the amplitude of $\psi_s$ is not observed near the 
central naked singularity and Cauchy 
horizon associated with it. 

\begin{figure}[htbp]
  \begin{center}
    \leavevmode
    \epsfysize=300pt\epsfbox{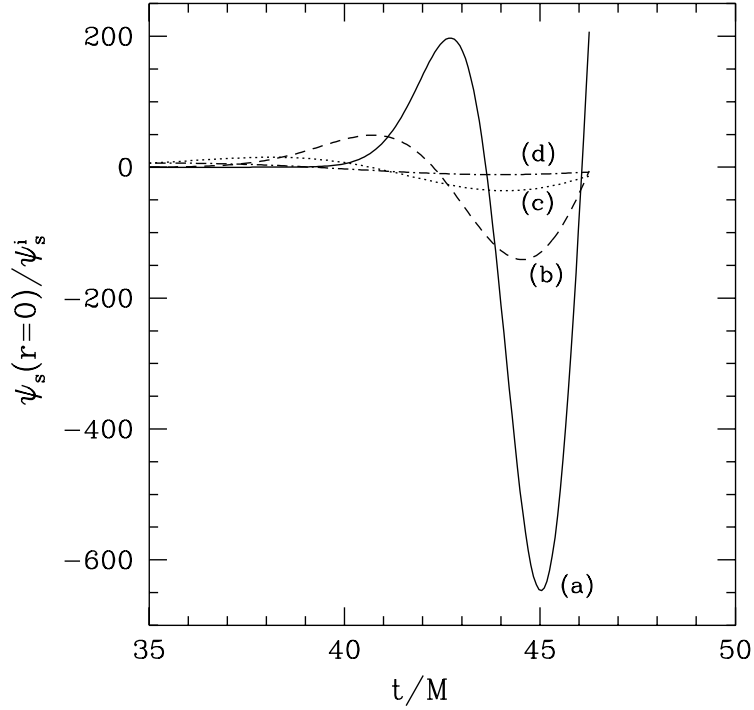}    
    \caption{Plots of the gauge-invariant variable, $\psi_s$, at the 
    center, $r=0$, 
    in the LTB space-time for various widths of initial wave packets. The wave 
    packets are put at $r=0.88r_b$ on the initial null surface. The widths
    of initial wave packets are varied from $0.03r_b$ to
    $0.12r_b$. The solid line (a) corresponds to the wave form of 
    the wave packet with the initial width $\sigma=0.03r_b$. 
    The broken line (b)
    is that of the initial width $\sigma=0.05r_b$ while the 
    dotted line (c) corresponds to that of the initial 
    width $\sigma=0.08r_b$. The broken dotted line 
    (d) is that of the initial width $\sigma=0.12r_b$.}
    \label{fig:sigma1}
  \end{center}
\end{figure}

\begin{figure}[htbp]
  \begin{center}
    \leavevmode
    \epsfysize=300pt\epsfbox{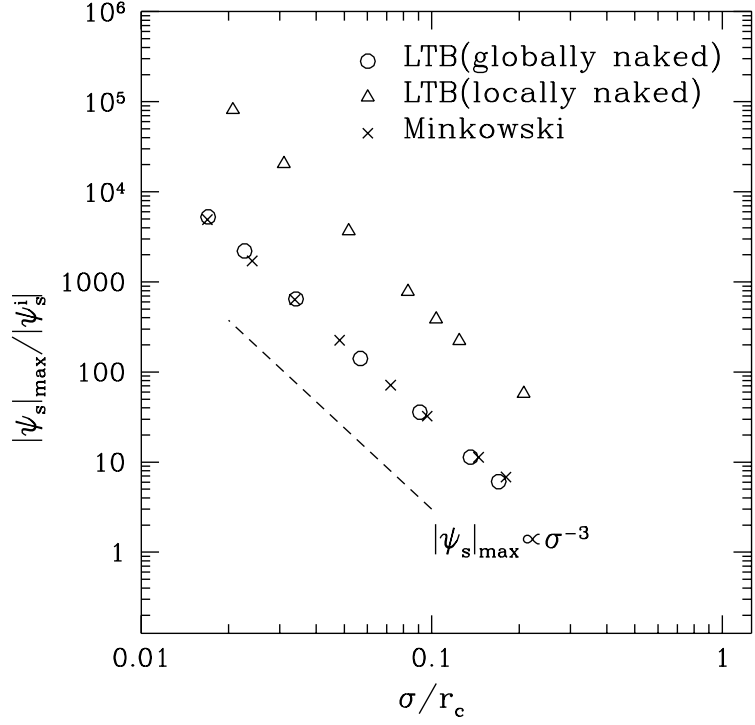}    
    \caption{The relation between the widths of initial wave packets and
    the maximal values of $|\psi_s|$ at the 
    center, $r=0$. The 
    results for the case of the LTB space-time with globally naked
    singularity are marked by open circles. The results of the locally
    naked case are marked by the triangles.  
    The results of the Minkowski space-time are marked by cross
    marks. The broken line denotes the relation, $|\psi_s|_{max}\propto
    \sigma^{-3}$.}
    \label{fig:sigma2}
  \end{center}
\end{figure}

Next we show the dependence of $\psi_s$ at the center 
on the width of the initial wave packet. 
Fig.\ref{fig:sigma1} depicts 
$\psi_s$ at the center for various widths of packets in Case 2. 
It is found that the amplitude of $\psi_s$
with smaller initial width becomes larger at the center. 
The relation between the width, $\sigma$, and 
the maximal value of $|\psi_s|$ at the center is shown in 
Fig.\ref{fig:sigma2}. 
We find that there is the following power-law relation 
\begin{equation}
  \label{power}
  |\psi_s|_{r=0} \propto \sigma ^{-3}.
\end{equation}

We also observe the time dependence of
$\psi_s$ along the line of a constant circumferential radius outside
the dust cloud.
Since we would like to see the effect 
of the central naked singularity on $\psi_{s}$, 
we consider the globally naked case only. 
We set up an initial wave packet of $\sigma=0.05r_{b}$ at
$R=100M$ on the initial null hypersurface which does not 
include the space-time point $(t,r)=(0,0)$ but 
is chosen so that the wave packet will reach the neighborhood 
of the central naked singularity. 

\begin{figure}[htbp]
  \begin{center}
    \leavevmode
    \begin{tabular}{c}
      \subfigure[Case 1,2,3]{\epsfysize=150pt\epsfbox{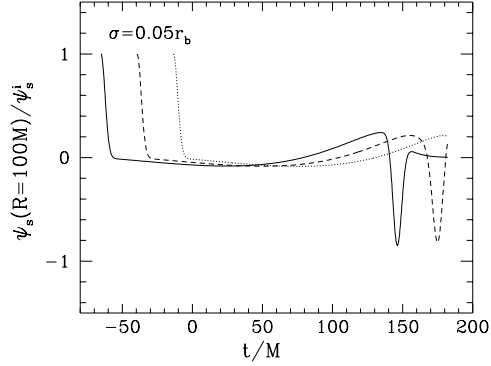}} \\
      \subfigure[Case 2]{\epsfysize=150pt\epsfbox{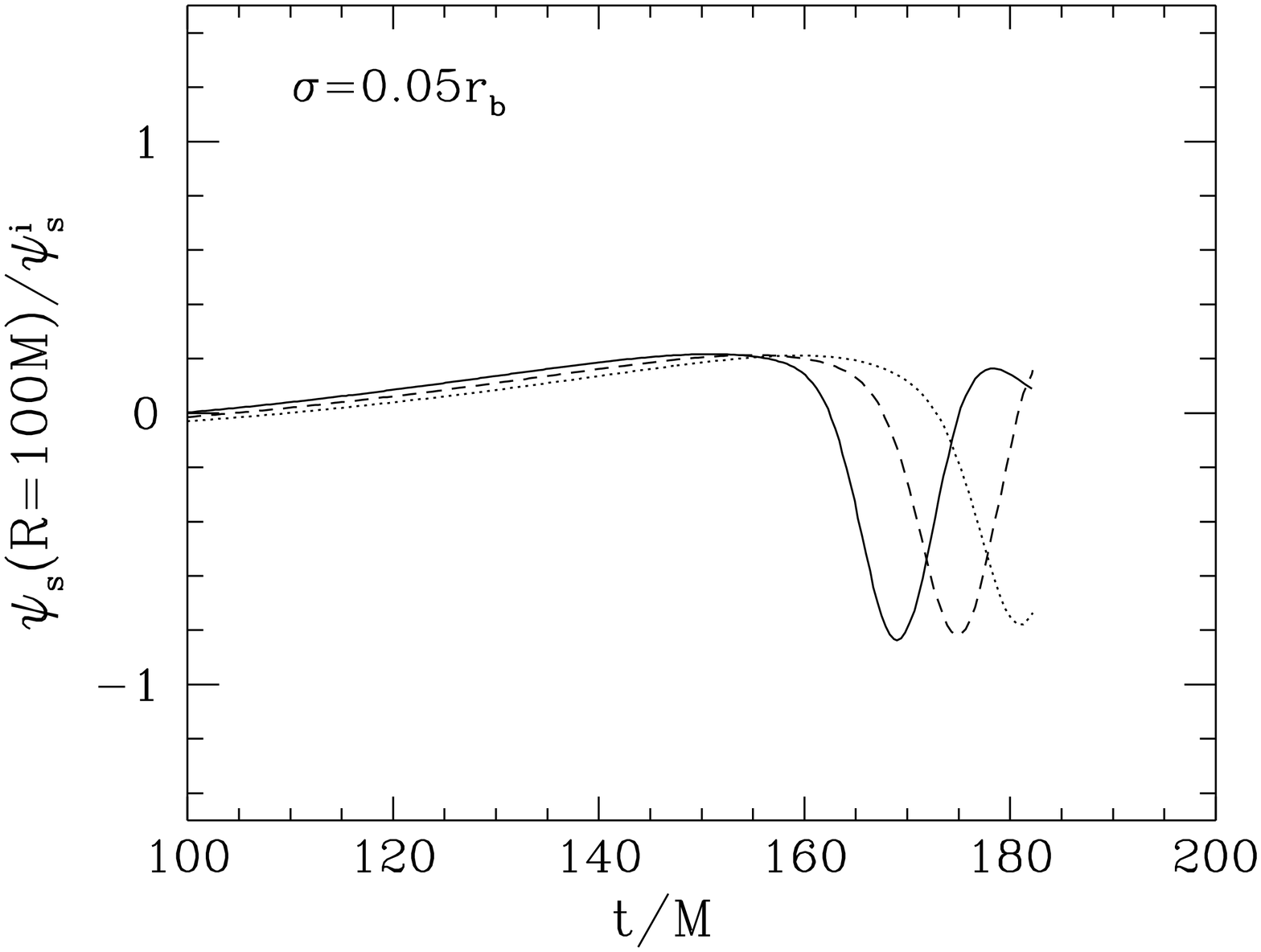}} \\
      \subfigure[$\sigma = 0.02-0.5r_b$ in Case2]{\epsfysize=150pt\epsfbox{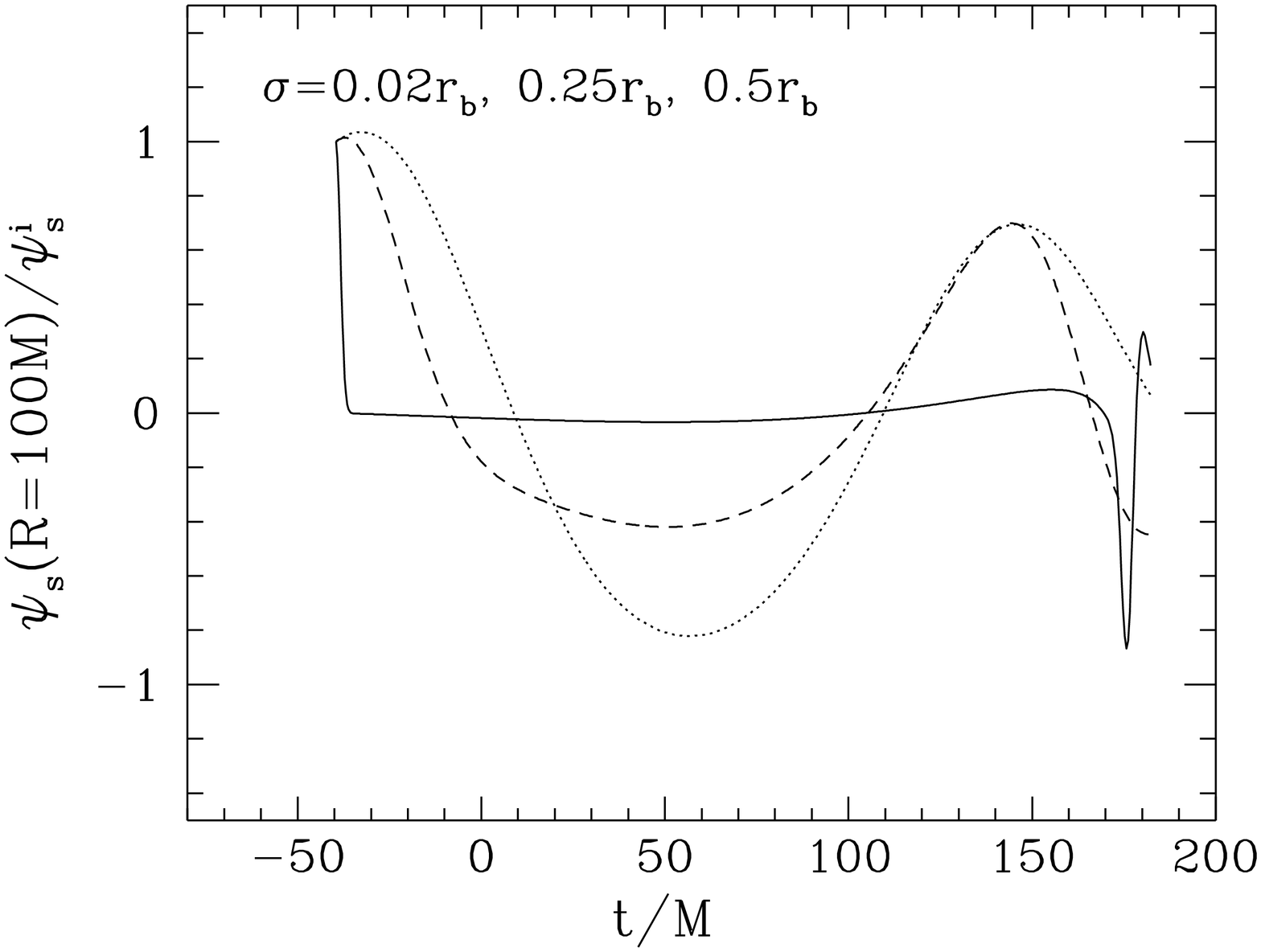}} 
    \end{tabular}
    \caption{Plots of $\psi_s$ with an initial width $\sigma=0.05r_b$
      at $R=100M$ as a function of time, $t$, in the LTB
    space-time. In (a), the solid line shows the result 
    for the case with an 
    initial time when we put the wave packet on the initial surface 
    $t_i/M=-65.310$ (Case1), the broken line is for $t_i/M=-38.529$ 
    (Case2), the dotted line is for $t_i/M=-13.610$ (Case3). 
    (b) depicts the
    details of the Case2. The solid line shows the plot of $\psi_s$ with
    the initial time, $t_i/M=-34.336$, the broken line is for $t_i/M=-38.529$, 
    the dotted
    line is for $t_i/M=-44.677$. We find no diverging tendency  
    of the gauge-invariant $\psi_s$ when it approaches the Cauchy
    horizon. In (c), we vary the width of the initial wave packet in
    Case2. The solid line is a plot of $\sigma = 0.02r_b$, the 
    broken line is the case of the initial width $\sigma = 0.25r_b$, 
    the dotted line is that of $\sigma = 0.5r_b$.}
    \label{fig:100m}
  \end{center}
\end{figure}

The results are shown in Fig.\ref{fig:100m} in which 
$\psi_{s}$ at $R=100M$ is plotted 
as a function of  
$t$. Note that the point, $R=100M$, is located in the 
vacuum region which is the Schwarzschild space-time by 
Birkhoff's theorem. Hence the value of $t$ along the curve of $R=100M$ 
agrees with that of the usual static time coordinate 
of the Schwarzschild space-time. 

In Fig.\ref{fig:100m}(a), the solid line corresponds to  
Case 1 while the broken line is for Case 2. 
The dotted line denotes the result for Case 3.  
The left-hand peaks in Fig.\ref{fig:100m}(a) 
correspond to the initial incident waves. 
On the other hand, the right-hand peaks of Cases 1 and 2 
in this figure correspond to the scattered outgoing waves. 
In Case 3, the right-hand peak does not exist and 
this is because, in this case, almost 
all portions of the incident waves enter into 
the Cauchy horizon associated with the central naked singularity 
and hence it is impossible to follow numerically 
the scattered waves in the causal future of the central 
naked singularity. Fig.\ref{fig:100m}(b) shows detailed behavior of the 
scattered $\psi_s$ for Case 2.
It is a most important fact seen in these figures that 
the amplitude of the scattered waves is almost the same as that of
the initial incident waves in Cases 1 and 2. 

In order to investigate the effect of the wavelength 
of $\psi_{s}$, we perform the numerical integration for Case 2 
but with different initial widths of wave packets. 
The results of narrower ($\sigma =0.02r_b$) and broader
($\sigma = 0.25r_b$ and $0.5r_b$) widths than the case plotted in 
Fig.\ref{fig:100m}(a) and (b) are shown in Fig.\ref{fig:100m}(c). 
The narrower wave is similar to 
$\sigma=0.05r_b$ while the broader packets have different forms 
of scattered waves from the narrower one. 
However, in both cases, the amplitude of the scattered wave 
is not so different from the incident one. 

\subsection{Minkowski Case}
Here we investigate the behavior of $\psi_{s}$  
in the Minkowski space-time and compare it with the results of 
the LTB space-time obtained in the above in order to reveal 
the effects of the existence of the dust cloud 
and the central naked singularity on the dynamics of $\psi_{s}$. 
In the Minkowski case, since $R(t,r)=r$, Eq.(\ref{wave-eq}) becomes 
\begin{equation}
  \label{Min}
    \partial_{t}^{2}\psi_{s}- \partial_{r}^{2}\psi_{s}
  = \frac{6}{r}\partial_{r}\psi_{s}.
\end{equation}
The solution of this equation which is regular at $r=0$ 
is obtained in the form
\begin{equation}
  \label{gen-sol}
  \psi_s = 3\frac{f(t-r)-f(t+r)}{r^5}+3\frac{f^{(1)}(t-r)+f^{(1)}(t+r)}{r^4}+
\frac{f^{(2)}(t-r)-f^{(2)}(t+r)}{r^3},
\end{equation}
where $f(x)$ is an arbitrary function and $f^{(n)}(x)$ denotes 
the $n$-th order derivative of $f(x)$ with respect to $x$. 
We set the following initial wave packet on the null 
hypersurface, $t=r$, 
\begin{equation}
  \label{min-ini}
  \psi_s = \exp\left[-\frac{(r-r_c)^2}{2\sigma^2}\right]. 
\end{equation}
Using the above solution, we compare the evolution of wave forms 
in the Minkowski space-time with that in the LTB space-time. 
The initial wave packet in the LTB space-time has been given by 
Eq.(\ref{initial}) as a function of the coordinate radius, $r'$. 
However, note that $r'$ does not agree with 
the circumferential radius, $R$, in this case 
but in the Minkowski case, the coordinate radius, $r$, 
agrees with the circumferential radius, $R$. 
Since the circumferential radius, $R$, is tightly connected with 
the behavior of the amplitude of the wave, 
we should set the same initial data with respect to 
$R$ both for the LTB and Minkowski cases.  
Hence first we plot the initial wave packet (\ref{initial}) 
as a function of $R/M$ on the initial null hypersurface 
and then the values of $\sigma$ and $r_c$ in Eq.(\ref{min-ini}) 
are adjusted so that the initial wave form fits well with that 
of the LTB case. 

\begin{figure}[htbp]
  \begin{center}
    \leavevmode
      \subfigure[Case 1]{\epsfysize=220pt\epsfbox{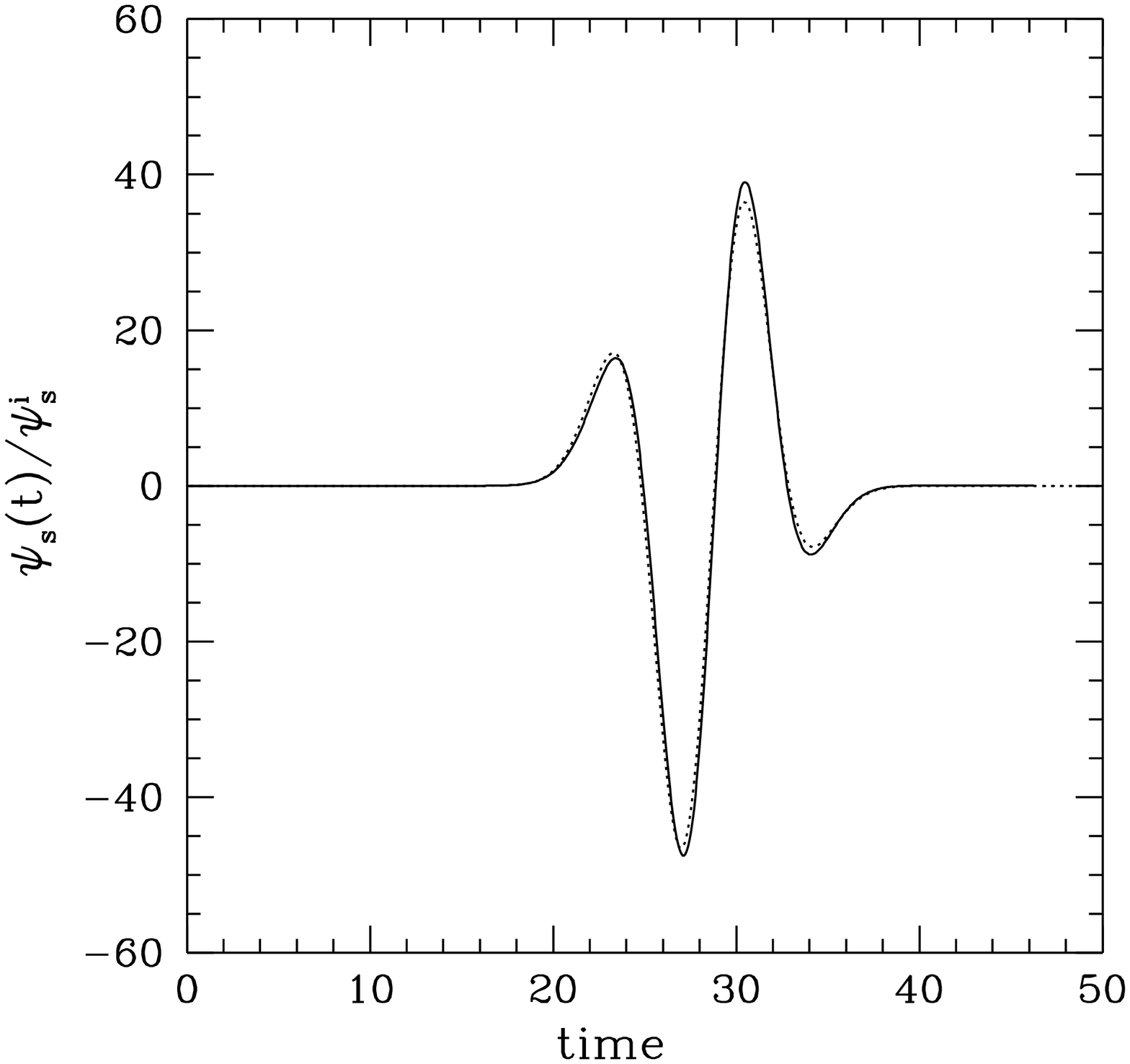}} 
      \subfigure[Case 2]{\epsfysize=220pt\epsfbox{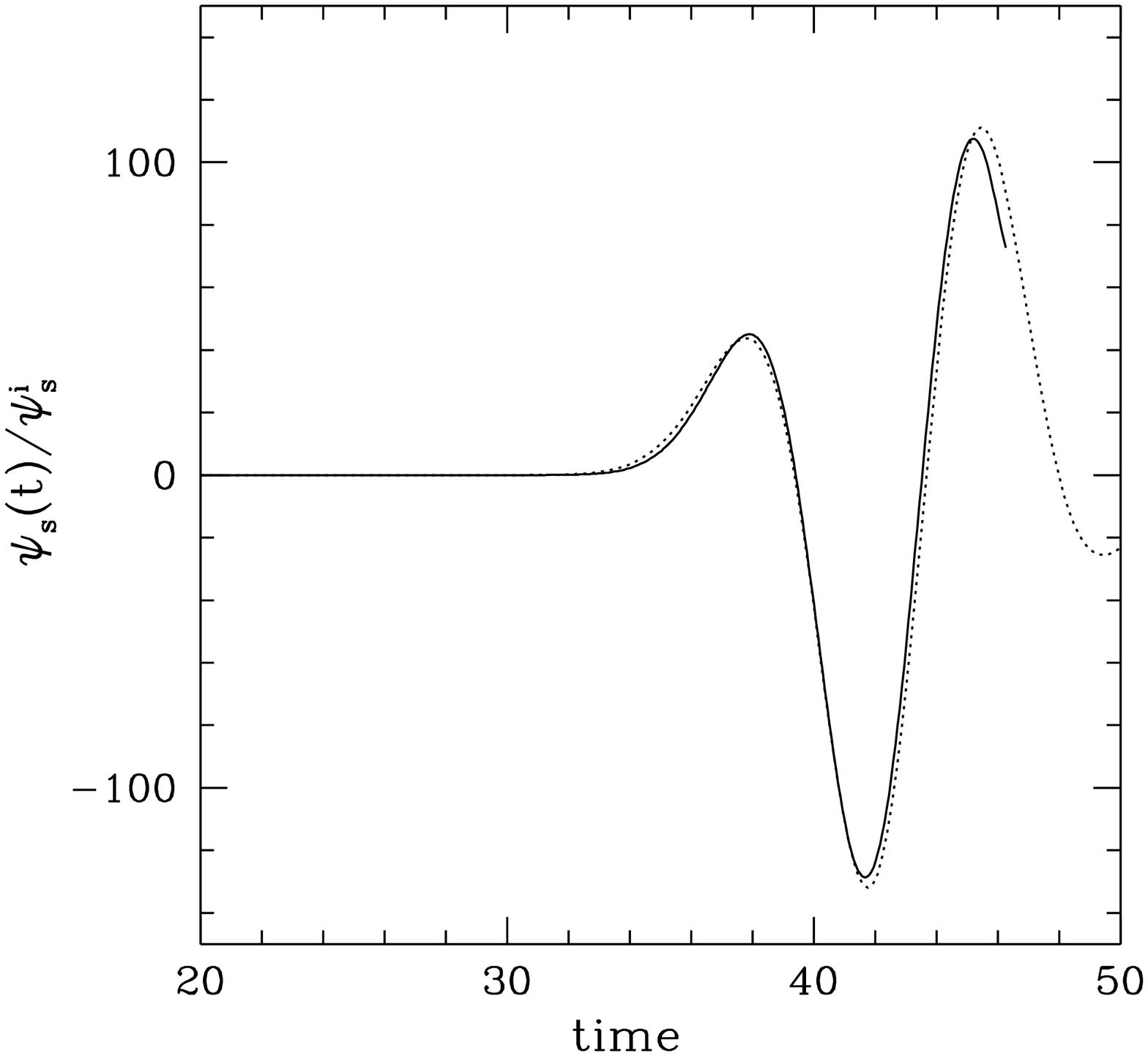}} 
    \caption{Results of the comparison of the wave forms at the
      center. In (a), the solid line shows the LTB case, as a function
      of $t/M$, that is the same as the 
      Case 1 in Fig.\protect\ref{fig:center1}(a). The dotted line shows the 
      corresponding one of the Minkowsi case, as a function of $t$,
      where $\sigma=1.18$ and 
      $r_c=13.9$. In (b), the LTB case is the solid line and Case 2 in
      Fig.\protect\ref{fig:center1}(a). The Minkowski case is the dotted
      line where $\sigma=1.25$ and $r_c=21.4$.}    
    \label{fig:compare1}
  \end{center}
\end{figure}

First we consider the evolution of $\psi_{s}$ at the center.  
Using Eqs.(\ref{gen-sol}) and (\ref{min-ini}), 
we obtain $\psi_s$ at the center in the form 
\begin{eqnarray}
  \label{min-r=0}
  \psi_s(t,0) &=&
  \left[1-\frac{1}{\sigma^2}\left(\frac{t}{2}-r_c\right)\frac{t}{2}-
  \frac{1}{4\sigma^2}
  \left(\frac{t}{2}\right)^2+\frac{1}{4\sigma^4}\left(
  \frac{t}{2}-r_c\right)^2 
  \left(\frac{t}{2}\right)^2\right.\nonumber\\
 & & \left.
 +\frac{1}{20\sigma^4}\left(\frac{t}{2}-r_c\right)
  \left(\frac{t}{2}\right)^3-\frac{1}{60\sigma^6}
  \left(\frac{t}{2}-r_c\right)^3
  \left(\frac{t}{2}\right)^3\right]\exp\left[-\frac{1}{2\sigma^2}
  \left(\frac{t}{2}-r_c\right)^2
  \right]. 
\end{eqnarray}
The parameters, $\sigma$ and $r_{c}$, in Eq.(\ref{min-ini}) 
are chosen so that the initial wave packets fit well with those 
of the Cases 1 and 2 of Fig\ref{fig:center1}. 
The results are given in Fig.\ref{fig:compare1}(a) and (b), 
respectively, and in this figure, 
we also plot the results for the corresponding cases of the 
LTB space-time. 
It should be noted that there is scarcely any difference 
between the wave forms of the Minkowski and LTB cases.  

\begin{figure}[htbp]
  \begin{center}
    \leavevmode
      \epsfysize=250pt\epsfbox{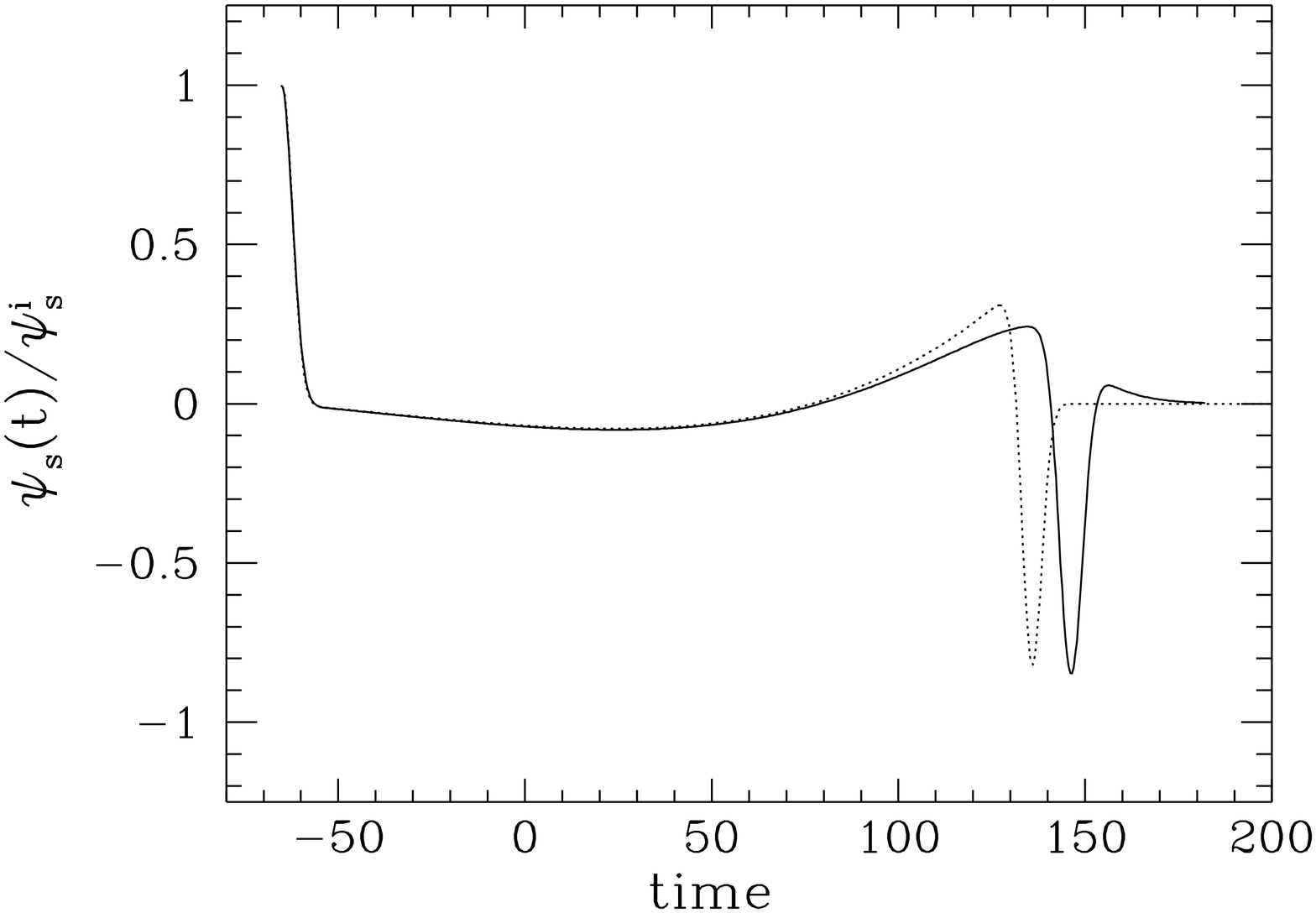}    
    \caption{Wave forms along the
      constant circumferential radius both for the Minkowski and LTB 
      cases are plotted. 
      The solid line shows the LTB case,
      as a function of $t/M$, that is identical with Case 1 in 
      Fig.\protect\ref{fig:100m} (a). The
      dotted line shows the corresponding one of the Minkowski case, as a 
      function of $t$, where
      $\sigma=1.33$ and $r_c=100$.}
    \label{fig:compare2}
  \end{center}
\end{figure}

Next we consider the behavior of $\psi_{s}$ at a finite 
circumferential radius which agrees with the 
numerical value of $R=100M$ in the LTB case.  
Here the wave form is obtained numerically by the same 
procedure as in the LTB case. 
The result is shown in Fig.\ref{fig:compare2}. 
We also plot the corresponding case of the LTB space-time 
in the same figure. 
We find that there is a little difference of the 
phase between the Minkowski and LTB cases. 
However, the behavior of $\psi_{s}$ in 
the Minkowski case is basically the same as that in the LTB case. 
The effect due to the dust cloud and 
the existence of the central naked singularity on the propagation 
of $\psi_{s}$ is rather small. 

We consider the relation between the 
maximum value of $|\psi_{s}|$ observed at the center 
and the width, $\sigma$, of an initial wave packet in the Minkowski 
space-time. This relation is obtained 
from Eq.(\ref{min-r=0}). The results are
also shown in Fig.\ref{fig:sigma2}. The power-law relation Eq.(\ref{power}) 
is also valid in the Minkowski case. From Eq.(\ref{gen-sol}), $\psi_s$ 
is approximately proportion to
$1/r^3$ except for the region of $r\alt\sigma$ around 
the center. 
If the initial amplitude of the wave packet has a value 
$\psi_i$ at $r=r_c$,
then the value of $\psi_s$ at $r=\sigma$ is roughly estimated as 
$\psi_i\times(\sigma/r_c)^{-3}$.
This will be the reason why the relation (\ref{power}) holds
in the Minkowski space-time. As we have discussed above, $\psi_s$
behaves outside the Cauchy horizon of the LTB space-time  in 
approximately the same manner as in the Minkowski space-time. Therefore it 
would be also 
the reason why the relation (\ref{power}) holds in the LTB space-time. 

As a result, we conclude that 
even in the neighborhood of the central naked singularity and 
of the Cauchy horizon associated with it, 
the metric perturbation, $\psi_{s}$, does not show any peculiar behavior. 
However, we should note that $\psi_{s}$ 
does not vanish in the neighborhood of the central naked 
singularity although it is well-behaved. 
Therefore, the formation process of the 
central naked singularity is {\it marginally stable} 
against the odd-parity metric perturbations. 

\section{CONCLUSIONS}
\label{sec:conclusions}
We have investigated the behavior of the odd-parity linear
perturbations in the LTB space-time. 
We have derived the wave equation for the
gauge-invariant variable, $\psi_s$. 
From the analysis of the regularity for $\psi_s$ and the perturbations 
of the Riemann tensor, 
only the quadrupole mode, $l=2$, of $\psi_s$ and of the
magnetic part of the Weyl tensor does not vanish at 
the symmetric center of the background LTB space-time, 
where a naked singularity appears in the course of the gravitational 
collapse of the dust cloud. Therefore
this quadrupole mode is the most important for the stability analysis of 
naked singularity formation in the LTB space-time. 
Then we have performed numerical
experiments on how a Gaussian-shaped incident wave packet behaves 
under this wave equation for the $l=2$ mode without matter perturbations. 
From those numerical experiments, we have obtained the following results. 
When this wave packet approaches the center, its amplitude 
becomes larger but finite. 
The amplitude at the center depends on the width of
the initial wave packet according to a power law. 
On the other hand, when the incident wave packet initially located outside 
the dust cloud returns back to the same circumferential radius as the initial 
one, the amplitude of the returned wave is almost equal to that of the 
incident one.  

In order to reveal the characteristic effects of the LTB space-time 
on the behavior of $\psi_{s}$, we have also investigated $\psi_{s}$ 
in the Minkowski space-time. Then we have found that 
in the outside of the Cauchy horizon associated with the central 
naked singularity, the behavior of $\psi_{s}$ in the LTB space-time 
seems to be not so different from that in the Minkowski space-time 
at least except for the extreme neighborhood of the naked singularity. 
Therefore the power-law dependence in the LTB space-time described above
is basically realized by the analytical discussion about the case of 
the Minkowski space-time.
Further the propagation effect due to the existence of the dust cloud 
and the occurrence of the central naked singularity is rather small. 
In other words, there is no peculiar behavior of $\psi_{s}$ 
even in the neighborhood of the central naked singularity. 
However, it should be noted that the odd-parity metric perturbation 
does not vanish in the neighborhood of the central 
naked singularity and Cauchy horizon associated with it. 
As a result, we conclude
that the central naked singularity formation in the LTB space-time is
`marginally' stable against the odd-parity metric perturbations.

We note that our analyses are not sufficient to determine the stability of
the naked singularity formation in the LTB space-time. 
There remain some problems to complete the analysis.
The first problem is to take account of 
odd-parity matter perturbations.
We are now investigating this problem.
The second is to consider the even-parity mode in which the 
metric and matter perturbations are essentially coupled with each other. 
This problem will be analyzed in future. 
In this paper, we have dealt with the marginally bound
case. For the case that is not marginally bound collapse, 
the condition of the appearance
of the central naked singularity is slightly different from the above
case\cite{Singh-Joshi,Jhingan} and hence there is a possibility 
that the behavior of $\psi_{s}$ in this case 
is different from the marginally bound one. 
This case is now under investigation. 

\section*{ACKNOWLEDGMENTS}
We would like to thank T. Nakamura, M. Sasaki, M. Shibata, 
T. Tanaka, M. Siino and T. Chiba 
for helpful and useful discussions, and N. Sugiyama and S. A. Hayward
for a careful reading of the manuscript. We are also
grateful to H. Sato and colleagues in the theoretical astrophysics group
in Kyoto University for useful comments and encouragement.

\appendix
\section{ANALYTIC SOLUTION OF THE DIPOLE MODE}
\label{sec:Dipole}
We present an analytic solution of the wave equations for the $l=1$
mode. Substituting $l=1$ into Eqs.(\ref{wave2}) and (\ref{wave3}), we get
\begin{eqnarray}
  \label{l=1-G-S-9}
  \partial_t\left(R^4\psi_s\right) &=& 0, \\
  \label{l=1-sita}
  \partial_r\left[R^4\psi_s-16\pi J(r)\right] &=& 0.
\end{eqnarray}
Eq.(\ref{l=1-sita}) is easily integrated and we obtain 
\begin{eqnarray}
  \label{sita}
  R^4\psi_s-16\pi J(r) &=& c(t),
\end{eqnarray}
where $c(t)$ is an arbitrary function of $t$. 
Substituting this equation into (\ref{l=1-G-S-9}), we obtain 
\begin{equation}
  \label{c(t)}
  \partial_tc(t) = 0.
\end{equation}
Hence the function, $c(t)$, is temporally constant. From 
Eq.(\ref{kakuundouryou}) and the regularity conditions for $R$, 
$L_0$ and $\psi_s$, the left hand side of Eq.(\ref{sita}) 
vanishes at the center. 
Therefore the function, $c(t)$, should vanish identically 
and the solution 
for the dipole mode is obtained in the form
\begin{equation}
  \label{Ks-l=1}
\psi_s =\frac{16\pi}{R^4}J(r).
\end{equation}
 From the regularity condition, (\ref{r-0-psis1}), 
$J(r)/R^{4}$, is identically zero at the regular center and hence 
this mode does not affect the formation of the central naked 
singularity.

\newpage
\newpage
\newpage
\newpage
\newpage
\end{document}